\documentclass[12pt]{elsarticle}
\pdfoutput=1
\usepackage{graphicx}
\usepackage[lined,algonl,boxed]{algorithm2e}
\usepackage{amssymb,amsthm}

\usepackage{bm}
\usepackage{asymptote}
\usepackage{lineno}
\usepackage{hyperref}

\usepackage{url}

\openin\AsyTestStream=cad-1.pre
\ifeof\AsyTestStream
\else
  \input cad-1.pre
\fi
\closein\AsyTestStream%

\openin\AsyTestStream=cad-2.pre
\ifeof\AsyTestStream
\else
  \input cad-2.pre
\fi
\closein\AsyTestStream%

\openin\AsyTestStream=cad-3.pre
\ifeof\AsyTestStream
\else
  \input cad-3.pre
\fi
\closein\AsyTestStream%

\openin\AsyTestStream=cad-4.pre
\ifeof\AsyTestStream
\else
  \input cad-4.pre
\fi
\closein\AsyTestStream%

\openin\AsyTestStream=cad-5.pre
\ifeof\AsyTestStream
\else
  \input cad-5.pre
\fi
\closein\AsyTestStream%

\openin\AsyTestStream=cad-6.pre
\ifeof\AsyTestStream
\else
  \input cad-6.pre
\fi
\closein\AsyTestStream%

\openin\AsyTestStream=cad-7.pre
\ifeof\AsyTestStream
\else
  \input cad-7.pre
\fi
\closein\AsyTestStream%

\openin\AsyTestStream=cad-8.pre
\ifeof\AsyTestStream
\else
  \input cad-8.pre
\fi
\closein\AsyTestStream%

\openin\AsyTestStream=cad-9.pre
\ifeof\AsyTestStream
\else
  \input cad-9.pre
\fi
\closein\AsyTestStream%

\openin\AsyTestStream=cad-10.pre
\ifeof\AsyTestStream
\else
  \input cad-10.pre
\fi
\closein\AsyTestStream%

\openin\AsyTestStream=cad-11.pre
\ifeof\AsyTestStream
\else
  \input cad-11.pre
\fi
\closein\AsyTestStream%

\openin\AsyTestStream=cad-12.pre
\ifeof\AsyTestStream
\else
  \input cad-12.pre
\fi
\closein\AsyTestStream%

\openin\AsyTestStream=cad-13.pre
\ifeof\AsyTestStream
\else
  \input cad-13.pre
\fi
\closein\AsyTestStream%

\openin\AsyTestStream=cad-14.pre
\ifeof\AsyTestStream
\else
  \input cad-14.pre
\fi
\closein\AsyTestStream%

\openin\AsyTestStream=cad-15.pre
\ifeof\AsyTestStream
\else
  \input cad-15.pre
\fi
\closein\AsyTestStream%

\openin\AsyTestStream=cad-16.pre
\ifeof\AsyTestStream
\else
  \input cad-16.pre
\fi
\closein\AsyTestStream%

\openin\AsyTestStream=cad-17.pre
\ifeof\AsyTestStream
\else
  \input cad-17.pre
\fi
\closein\AsyTestStream%

\openin\AsyTestStream=cad-18.pre
\ifeof\AsyTestStream
\else
  \input cad-18.pre
\fi
\closein\AsyTestStream%

\journal{Computer Aided Design}

\newtheorem{theorem}{Theorem}
\begin{asydef}
import three;
settings.render=4;
viewportmargin=(2,2);
\end{asydef}

\def\acro#1{#1}
\def\Asymptote{{\sc Asymptote}}
\def\PS{{Post\-Script}}
\def\OpenGL{{\sc OpenGL}}
\def\PDF{\acro{PDF}}
\def\PRC{\acro{PRC}}
\def\SVG{\acro{SVG}}
\def\EMF{\acro{EMF}}
\def\It#1{{\it #1}}
\def\Jacobian#1#2{\frac{\partial(#1)}{\partial(#2)}}
\def\v{\bm}
\def\grad{\v\nabla}
\def\cross{{\v\times}}

\def\vP{{\v P}}
\def\vU{{\v U}}
\def\vV{{\v V}}

\def\va{{\v a}}
\def\vb{{\v b}}

\def\vp{{\v p}}
\def\vq{{\v q}}

\def\R{\mathbb R}
\def\Box{\mathop{\rm box}\nolimits}
\def\degrees{\circ}
\def\etal{{\it et al.}}

\usepackage{mflogo}

\begin{document}

\def\today{May 17, 2010}

\begin{frontmatter}

\title{Surface Parametrization of Nonsimply Connected Planar B\'ezier Regions} 

\author{Orest Shardt}
\address{Department of Chemical and Materials Engineering, 
University of Alberta, Edmonton, Alberta T6G 2V4, Canada}
\author{John C. Bowman\corref{cor1}}
\address{Department of Mathematical and Statistical Sciences,
University of Alberta, Edmonton, Alberta T6G 2G1, Canada}
\cortext[cor1]{Corresponding author}
\ead[url]{http://www.math.ualberta.ca/$\sim$bowman}

\begin{abstract}
A technique is described for constructing three-dimensional vector graphics
representations of planar regions bounded by cubic B\'ezier curves, such as
smooth glyphs. It relies on a novel algorithm for compactly partitioning
planar B\'ezier regions into nondegenerate Coons patches. New optimizations are
also described for B\'ezier inside--outside tests and the computation of
global bounds of directionally monotonic functions over a B\'ezier surface
(such as its bounding box or optimal field-of-view angle). These algorithms
underlie the three-dimensional illustration and typography features of
the \TeX-aware vector graphics language \Asymptote.
\end{abstract}





\end{frontmatter}

\linenumbers

\bibliographystyle{elsarticle-harv}

\section{Introduction}

Recent methods for lifting smooth two-dimensional (2D) font data into three
dimensions (3D) have focused on rendering algorithms for the Graphics Processing
Unit (GPU) \cite{Loop05}. However, scientific visualization often requires
3D vector graphics descriptions of surfaces constructed from smooth font
data. For example, while current CAD formats, such
as the \PDF-embeddable \It{Product Representation Compact} 
(PRC, \It{pr\'ecis} in French) \cite{PRCFormat08} format, allow one to
embed text annotations, they do not allow text to be manipulated as a 3D
entity. Moreover, annotations can only handle simple text; they are not
suitable for publication-quality mathematical typesetting.

In this work, we present a method for representing arbitrary planar
regions, including text, as 3D surfaces. A significant advantage of this
representation is consistency: text can then be rendered like any other 3D
object.  This gives one complete control over the typesetting process, such
as kerning details, and the ability to manipulate text arbitrarily (e.g.\ by
transformation or extrusion) in a compact resolution-independent vector
form. In contrast, rendering and mesh-generation approaches destroy
the smoothness of the original 2D font data.

In focusing on the generation of 3D surfaces from 2D planar data, the emphasis
of this work is not on 3D rendering but rather on the underlying procedures for
generating vector descriptions of 3D geometrical objects. Vector
descriptions are particularly important for
online publishing, where no assumption can be made \It{a priori} about the
resolution that will be used to display an image.
As explained in Section~\ref{beziervsnurbs}, we focus on surfaces based on
polynomial parametrizations rather than nonuniform rational B-splines
(\acro{NURBS}) \cite{Farin92,Piegl97}.
In Section~\ref{beziertriangulation} we describe a method for splitting
an arbitrary planar region bounded by one or more B\'ezier curves into
nondegenerate B\'ezier patches.
This algorithm relies on the optimized B\'ezier inside--outside test
described in Section~\ref{insidedness}.
The implementation of these algorithms in the vector graphics
language \Asymptote, along with the optimized 3D sizing algorithms
presented in Section~\ref{monotonic}, is discussed in Section~\ref{Asymptote}.

Using a compact vector format instead of a large number of polygons 
to represent manifolds has the advantage of  
reduced data representation (essential for the storage and transmission of 3D
scenes) and the possibility, using relatively few control points, 
of exact or nearly exact geometrical descriptions of mathematical surfaces.
For example, in Appendix~\ref{solids} we show that a sphere can be
represented to $0.05\%$ accuracy with just eight cubic B\'ezier surface patches.

\section{B\'ezier vs. NURBS Parametrizations}\label{beziervsnurbs}
The atomic graphical objects in \PS\ and \PDF, B\'ezier curves and surfaces,
are composed of piecewise cubic polynomial \It{segments} and tensor
product \It{patches}, respectively. A segment
$\v\gamma(t)=\sum_{i=0}^3 B_i(t) \vP_i$ has four control points~$\vP_i$,
whereas a surface patch is defined by sixteen control points~$\vP_{ij}$:
$$
\v\sigma(u,v)=(x(u,v),y(u,v))=\sum_{i,j=0}^3 B_i(u)B_j(v) \vP_{ij}.
$$
Here $B_i(u)={3\choose i} u^i(1-u)^{3-i}$ is the $i$th cubic Bernstein
polynomial. Just as a B\'ezier curve passes
through its two end control points, a B\'ezier surface necessarily passes
through its four corner control points. These special control points are called
\It{nodes}.
It is convenient to define the {\it convex hull\/} of a
cubic B\'ezier segment or patch to be the convex hull (minimal enclosing
polygon or polyhedron) of its control points.
A \It{straight} segment is one in which the control
points are colinear and the derivative of the B\'ezier parametrization is
never zero (i.e.\ the control points are arranged in the same order as their
indices).

It is often desirable to project a 3D scene to a 2D vector graphics format
understood by a web browser or high-end printer.  Although \acro{NURBS} are
popular in computer-aided design \cite{Farin92} because of the additional
degrees of freedom introduced by weights and general knot vectors, these
benefits are tempered by both the lack of support for \acro{NURBS} in popular 2D
vector graphics formats (\PS, \PDF, \SVG, \EMF) and the algorithmic
simplifications afforded by specializing to a B\'ezier parametrization.  
B\'ezier curves are also commonly used to describe glyph outlines.
We therefore restrict our attention to (polynomial) B\'ezier curves and
surfaces (even though both \Asymptote\ and the 3D \PRC\ format support
\acro{NURBS}).

Unlike their B\'ezier counterparts, \acro{NURBS} are invariant
under perspective projection. This is only an issue if
projection is done before the rendering stage, as is necessary when
a 2D vector representation of a curve or surface is constructed solely from
the 2D projection of its control points.
It is therefore somewhat ironic that \acro{NURBS}
are much less widely implemented in 2D vector graphics formats than in 3D.
In 3D vector graphics applications, projection to 2D is always deferred
until rendering time, so that the invariance of \acro{NURBS} under
nonaffine projection is irrelevant.
While \acro{NURBS} provide exact parametrizations of
familiar conic sections and quadric surfaces,
nontrivial manifolds still need to be approximated as piecewise unions of
underlying exact primitives. We feel that the
implementational simplicity of basic B\'ezier operations (computing
subcurves and subsurfaces, points of tangency, normal vectors, bounding
boxes, intersection points, arc lengths, and arc times) offsets for many
practical applications the lower dimensionality of the
B\'ezier subspace. 

\section{Partitioning Curved 2D Regions}\label{beziertriangulation}
In 3D graphics, text is often displayed with bit-mapped images, textures, or
polygonal mesh approximations to smooth font character curves. To allow viewing
of smooth text at arbitrary magnifications and locations, a nonpolygonal 
surface that preserves the curvature of the boundary curves is required.
While it is easy to fill the outline of a smooth character in 2D, filling a 3D
planar surface requires more sophisticated methods. One approach involves using
surface filling algorithms for execution on GPUs \cite{Loop05}.
When a vector, rather than a rendered, image is desired, a preferable
alternative is to represent the text as a parametrized surface.

Methods based on common surface primitives in 3D modelling and
rendering can be used to describe planar regions.
One method trims the domain of a planar surface to the desired
shape \cite{Nishita90}. While that approach is feasible, given adequate software
support for trimming, this work describes a different approach, where
each symbol is represented as a set of planar B\'ezier patches.
We call this procedure \It{bezulation} since it
involves a process similar to the triangulation of a polygon
but uses cubic B\'ezier patches instead of triangles.
To generate a surface representing the region bounded by a set of
\It{simple closed} B\'ezier
curves (intersecting only at the end points), algorithms were developed for
(i) expressing a simply connected 2D region as a union of
B\'ezier patches and (ii) breaking up a nonsimply connected region into
simply connected regions. (Selfintersecting curves can be handled
by splitting at the intersection points.) These algorithms allow one to
express text surfaces conveniently as B\'ezier patches.

\begin{figure}[tp]
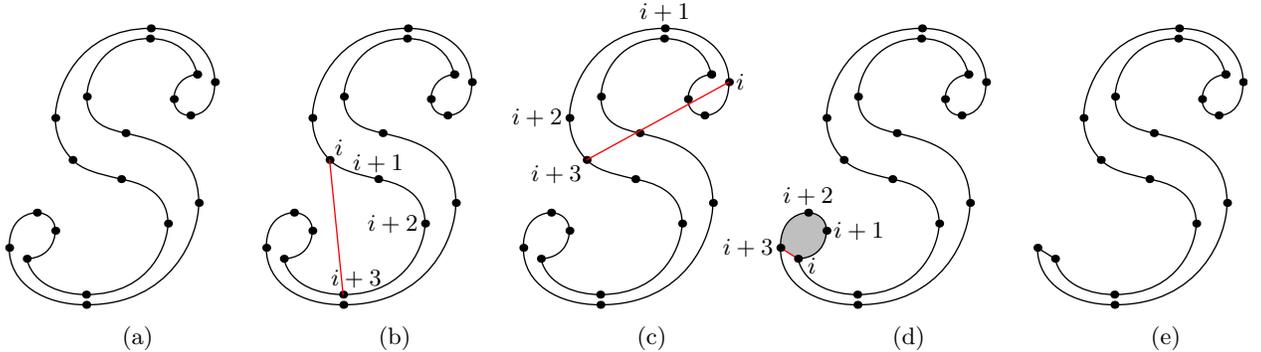

\begin{center}
\begin{asy}
include bezulateExample;
\end{asy}
\caption{The {\tt bezulate} algorithm. Starting with the original curve
(a), several possible connecting line segments (shown in red) between nodes
separated by $n=3$ or $n=2$ segments are tested. Connecting line segments are
rejected if they do not lie entirely inside the original curve. This occurs when
the midpoint is not inside the curve (b) or when the connecting line
segment intersects the curve more than twice (c). If a connecting line
segment passes both tests, the shaded section is separated (d) and the
algorithm continues with the remaining curve~(e).
}\label{bezulateExample}
\end{center}
\end{figure}

Bezulation of a simply connected planar region involves breaking
the region up into patches bounded by closed B\'ezier curves with four
or fewer segments. This is performed by the routine
{\tt bezulate} (cf.~Algorithm~\ref{bezulate}) using an adaptation of a
na\"ive triangulation algorithm, modified to handle curved edges,
as illustrated in Figure~\ref{bezulateExample}.

\SetAlCapSkip{3pt}

\begin{algorithm}[H]
  \SetKwData{Currentpath}{C}
  \SetKwData{Outputpath}{A}
  \SetKwData{Currentpathlength}{C.segments}
  \SetKwData{found}{found}
  \SetKw{Break}{break}
  \SetKw{And}{and}
  \SetKw{True}{true}
  \SetKw{False}{false}
  \KwIn{simple closed curve \Currentpath}
  \KwOut{array of closed curves \Outputpath}
  \While{\Currentpathlength $>$ 4}{
    found $\leftarrow$ \False;\\
    \For{$n$ = 3 \KwTo 2}{
      \For{$i$ = 0 to \Currentpathlength-1}{
          L $\leftarrow$ line segment between nodes $i$ and $i+n$ of \Currentpath\;
        \If{{\tt countIntersections}(\Currentpath,L) = 2 \And midpoint of $L$ is inside \Currentpath}{
          p $\leftarrow$ subpath of \Currentpath from node $i$ to $i+n$\;
          q $\leftarrow$ subpath of \Currentpath from node $i+n$ to $i+\Currentpathlength$\;
          \Outputpath.{\tt push}(p+$L$);\\
          \Currentpath $\leftarrow L$ $+$ q;\\
          found $\leftarrow$ \True;\\
          \Break;
        }
      }
      \If{\found}{\Break;}
    }
    \If{not \found}{
      refine \Currentpath by inserting an additional node at the parametric midpoint of each segment\;
    }
  }
\caption{{\tt bezulate} partitions a simply connected region.}\label{bezulate}
\end{algorithm}

A line segment lies within a closed curve when it intersects
the curve only at its endpoints and its midpoint lies strictly inside the curve.
If after checking all connecting line segments between nodes separated by
$n=3$ or $n=2$ segments, none of them lie entirely inside the shape, the
original curve is refined by dividing each segment of the curve at its
parametric midpoint. The bezulation process then continues with the refined
curve. This algorithm can be modified to subdivide more optimally, for
example, to avoid elongated patches that sometimes lead to rendering problems.

If the region is convex, Algorithm~\ref{bezulate} is easily seen to
terminate: all connecting line segments are admissible, and each patch
removal decreases the number of points in the curve.
Moreover, from the point of view of Algorithm~\ref{bezulate}, upon sufficient
subdivision a non-convex region eventually becomes indistinguishable from a
polygon, in which case the algorithm reduces to a straightforward polygonal
triangulation.

\subsection{Nonsimply Connected Regions}
Since the {\tt bezulate} algorithm requires simply connected
regions, nonsimply connected regions must be handled specially.
The ``holes'' in a nonsimply connected domain can be removed by partitioning
the domain into a set of simply connected regions,
each of which can then be bezulated.

For convenience we define a \It{top-level curve} to be a curve that is
not contained inside any other curve and an \It{outer (inner) curve} to be the
outer (inner) boundary of a filled region.
With these definitions, the glyph ``\%'' has two inner curves and two
top-level curves that are also outer curves. 

The algorithm proceeds as follows. First, to determine the topology of the
region, the curves are sorted according to their relative insidedness, as
determined by the nonzero winding number rule. Since the curves are assumed to
be simple, any point on an inner curve can be used to test
whether that curve is inside another curve. The result of this sorting is a
collection of top-level curves grouped with the curves they surround. Each
of these groups is treated independently.

\begin{figure}[tp]
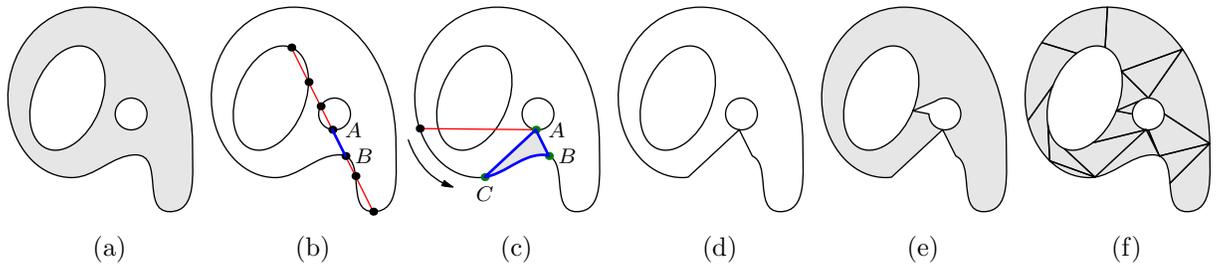

\begin{asy}
include mergeExample;
\end{asy}
\caption{Splitting of non-simply connected regions into simply connected 
  regions. Starting with a non-simply connected region~(a), the
  intersections between each curve and an arbitrary line segment from a
  point on an inner curve to the outer curve are found~(b).
  Consecutive intersections of this line segment, at points $A$ and $B$,
  on the inner and outer curves, respectively, identify a convenient
  location for extracting a region. One searches along the outer curve for
  a point $C$ such that the line segment $AC$ intersects the outer curve no
  more than once, intersects an inner curve only at $A$, and determines a
  region $ABC$ between the inner and outer curves that does not contain an inner
  curve. Once such a region is found (c), it is extracted (d). This
  extraction merges the inner curve with the outer curve. The process is
  repeated until all inner curves have been merged with the outer curve,
  leaving a simply connected region~(e) that can be split into B\'ezier
  surface patches. The resulting patches and extracted regions are
  shaded in~(f).}
\label{mergeExample}
\end{figure}

\begin{figure}[tp]
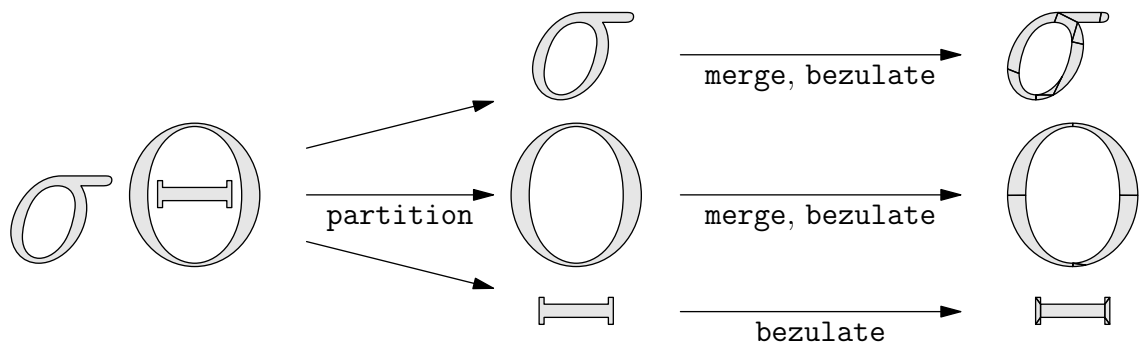

\begin{center}
\begin{asy}
include partitionExample;
\end{asy}
\caption{Illustration of the {\tt partition} algorithm. The five curves
that define the outlines of the Greek characters $\sigma$ and $\Theta$ are
passed in a single array to {\tt partition}.
}\label{partitionExample}
\end{center}
\end{figure}

Figure~\ref{partitionExample} illustrates the {\tt partition} routine
(cf.~Algorithm~\ref{partition}). 
Each group is examined recursively to identify regions bounded by
inner and outer curves. First, the inner curves
in the group are sorted topologically to find the inner
curves that are top-level curves with respect to the other inner curves. The
inner curves that are not top-level curves are processed with a recursive call
to {\tt partition}. The nonsimply connected region between the outer
(top-level) curve and the inner (top-level) curves is now split into simply
connected regions. This is illustrated in Figure~\ref{mergeExample}. The
intersections of the inner and outer curves with a line segment from a point on
an inner curve to a point on the outer curve are found (either {\it via\/}
subdivision or a numerically robust cubic root solver).
Consecutive intersections of this line segment, at points $A$ and $B$,
on the inner and outer curves, respectively, are selected. Let $t_B$
be the value of the parameter used to parameterize the outer curve at
$B$. Starting with $\Delta = 1$, $\Delta$ is halved until the line segment
$\overline{A C}$, where $C$ is the point on the outer curve at
$t_B+\Delta$, does not intersect the outer curve more than once, does not
intersect any inner curve (other than once at $A$), and the region bounded by
$\overline{A B}$, $\overline{A C}$, and $\stackrel{\frown}{B C}$
does not contain any inner curves. Once $\Delta$ and the point $C$ have been
found, the outer curve, less the segment between $B$ and $C$, is merged with
$\overline{B A}$, followed by the inner curve and then $\overline{A C}$.
The region bounded by $\overline{A B}$, $\overline{A C}$, and
$\stackrel{\frown}{B C}$ is a simply connected region. Additional simply
connected regions are found when the outer curve is merged with the other inner 
curves. Once the merging with all inner curves has been completed, the outer
curve becomes the boundary of the final simply connected region.

The recursive algorithm for partitioning nonsimply connected regions into
simply connected regions is summarized below. The function {\tt sort} returns
groups of top-level curves and the curves they contain. However, it is not
recursive; the inner curves are not sorted. The function 
{\tt merge} returns the simply connected regions formed from the single outer
curve and multiple inner curves that are supplied to it.

\begin{algorithm}[H]
  \SetKwData{InPathArray}{C}
  \SetKwData{OutPathArray}{A}
  \SetKwData{GG}{G}
  \SetKwData{InnerGroups}{innerGroups}
  \SetKwData{iG}{H}

  \KwIn{array of simple closed curves \InPathArray}
  \KwOut{array of closed curves \OutPathArray}

  \ForEach{group of nested curves \GG in {\tt sort}(\InPathArray)}{
    \InnerGroups $\leftarrow$ {\tt sort}(\GG.innerCurves)\;
    \ForEach{group of nested curves \iG in \InnerGroups}{
      \OutPathArray.{\tt push}({\tt partition}(\iG.innerCurves))\;
    }
    \OutPathArray.{\tt push}({\tt merge}(\GG.toplevel, top-level curves of all groups in \InnerGroups))\;
  }
  \Return \OutPathArray\;
\caption{{\tt partition} splits nonsimply connected regions
into simply connected regions. The pseudo-code functions {\tt sort} and 
{\tt merge} are described in the text.}\label{partition}
\end{algorithm}

\begin{figure}[tp]
\begin{center}
\begin{minipage}{0.49\linewidth}
\begin{center}
\begin{asy}[inline,width=\the\linewidth]
include label3solid;
\end{asy}
\caption{Application of the {\tt bezulate} and {\tt partition} algorithms
to lift the Gaussian integral to three dimensions.
}\label{label3solid}
\end{center}
\end{minipage}
\begin{minipage}{0.49\linewidth}
\begin{center}
\begin{asy}[inline,width=\the\linewidth]
include label3zoom;
\end{asy}
\caption{Zoomed view of Figure~\ref{label3solid} generated from the same
vector graphics data. The smooth boundaries of the characters 
emphasize the advantage of a 3D vector font description.}
\label{label3zoom}
\end{center}
\end{minipage}
\end{center}
\end{figure}

\begin{figure}[tp]
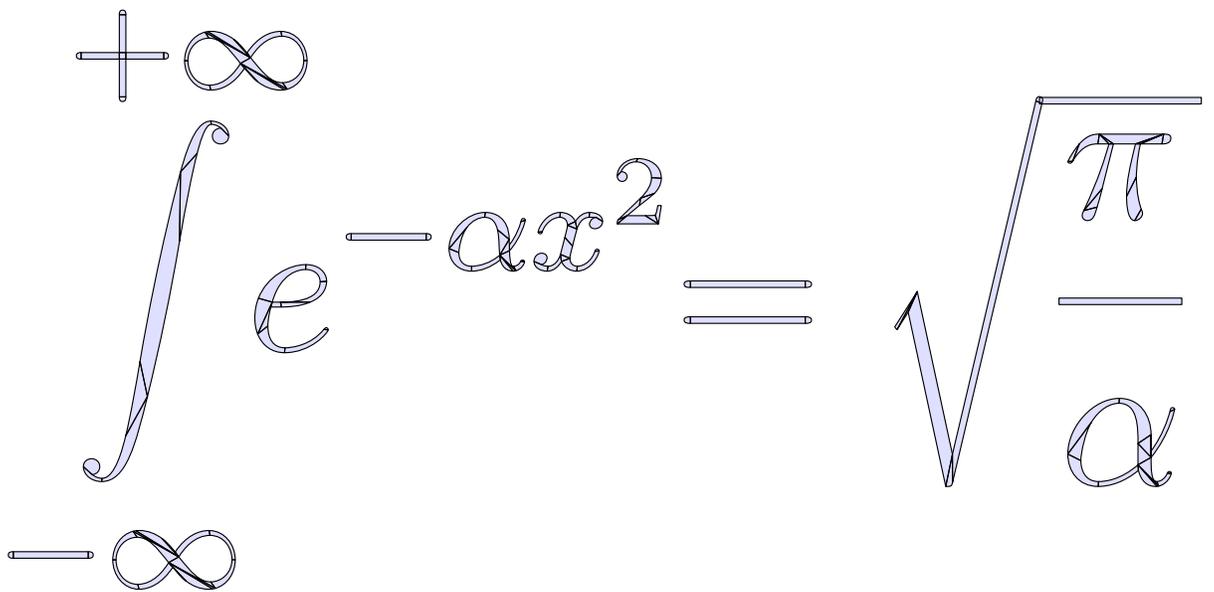

\begin{center}
\begin{asy}[inline]
include label3;
\end{asy}
\caption{Subpatch boundaries for Figure~\ref{label3solid} as determined by
  the {\tt bezulate} and {\tt partition} algorithms.
}\label{label3}
\end{center}
\end{figure}

The routines {\tt bezulate} and {\tt partition} were used
to typeset the \TeX\ equation 
$$
\int\limits_{-\infty}^{+\infty}\!\! e^{-\alpha x^2}\!\!=\sqrt{\frac{\pi}{\alpha}}
$$
in the interactive 3D diagram shown in Figure~\ref{label3solid}
and magnified, to emphasize the smooth font boundaries, in
Figure~\ref{label3zoom}.
The computed subpatch boundaries are indicated in Figure~\ref{label3}. 

Figure~\ref{unitsphere} in Appendix~\ref{solids} illustrates how
{\tt bezulate} is used in mathematical drawings to lift \TeX\ to three
dimensions. Referring to the interactive 3D \PDF\ version of
this article\footnote{See \url{http://asymptote.sourceforge.net/articles/}.}
one see that the labels in Figure~\ref{unitsphere} have been programmed to
rotate interactively so that they always face the camera; this feature,
implemented with {\tt Javascript}, is known as \It{billboard interaction}.

Developing B\'ezier versions of more sophisticated triangulation algorithms
would be an interesting future research project. The rendering technique of
Ref.~\cite{Loop05} could be modified to produce B\'ezier patches, but
this would produce more patches than {\tt bezulate}. For example, the ``e''
shown in Fig.~3 of Ref.~\cite{Loop05} corresponds to roughly twice as
many (4-segment) patches as the ten patches generated by {\tt bezulate} for
the ``$e$'' in Fig.~\ref{label3}.
Since our interest is in compact 3D vector representations, the objective
of this work is to minimize the number of generated patches. In contrast,
in real-time rendering, one aims to minimize the overall execution time.

\subsection{Nondegenerate Planar B\'ezier Patches}
The {\tt bezulate} algorithm described previously decomposes regions
bounded by closed curves (according to the nonzero winding
number rule) into subregions bounded by closed curves with four or fewer
segments. Further steps are required to turn these subregions into
nondegenerate B\'ezier patches.
First, if the interior angle between the incoming and outgoing
tangent directions at a node is greater than $180^\degrees$, the boundary
curve is split at this node by following the interior angle bisector
to the first intersection with the path.
This is done to guarantee that the patch normal
vectors at the nodes all point in the same direction. Next, curves with
less than four segments are supplemented with null segments (four identical
control points) to bring their total number of segments up to four.
A closed curve with four segments defines the twelve boundary control
points of a B\'ezier patch in the $x$--$y$ plane. The remaining four
interior control points
$\{\vP_{11},\vP_{12},\vP_{21},\vP_{22}\}$ are then chosen to satisfy the
Coons interpolation \cite{Coons64,Farin02,PDFReference08}
\begin{eqnarray}
\v\sigma(u,v)&=&\sum_{i=0}^3\left[(1-v)B_i(u)\vP_{i,0}+vB_i(u)\vP_{i,3}+
(1-u)B_i(v)\vP_{0,i}+uB_i(v)\vP_{3,i}\right]
\nonumber\cr
&&\qquad
-(1-u)(1-v)\vP_{0,0}-(1-u)v\vP_{0,3}-u(1-v)\vP_{3,0}-uv\vP_{3,3}.
\end{eqnarray}

\begin{figure}[tp]
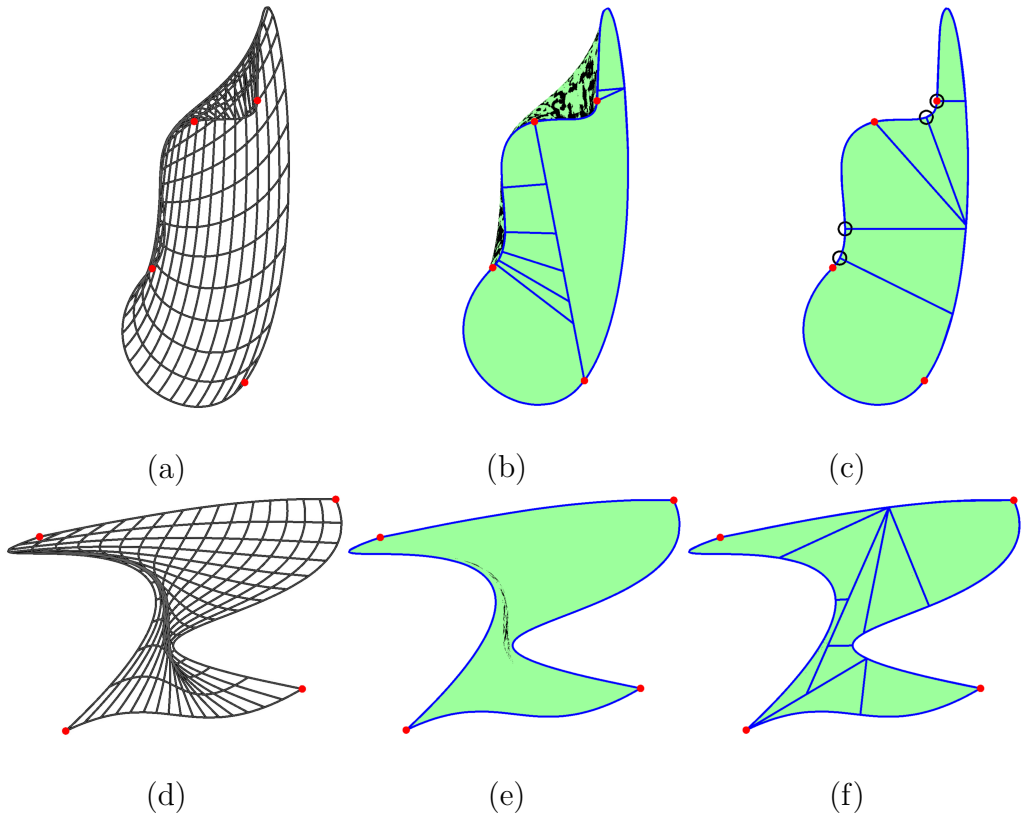

\begin{center}
\begin{minipage}{0.32\linewidth}
\begin{asy}[inline]
include overlapmesh;
\end{asy}
\begin{center}
(a)
\end{center}
\end{minipage}
\begin{minipage}{0.32\linewidth}
\begin{asy}[inline]
include overlapsurface;
\end{asy}
\begin{center}
(b)
\end{center}
\end{minipage}
\begin{minipage}{0.32\linewidth}
\begin{asy}[inline]
include overlapsurfacefixed;
\end{asy}
\begin{center}
(c)
\end{center}
\end{minipage}
\begin{minipage}{0.32\linewidth}
\begin{asy}[inline]
include degeneratemesh;
\end{asy}
\begin{center}
(d)
\end{center}
\end{minipage}
\begin{minipage}{0.32\linewidth}
\begin{asy}[inline]
include degeneratesurface;
\end{asy}
\begin{center}
(e)
\end{center}
\end{minipage}
\begin{minipage}{0.32\linewidth}
\begin{asy}[inline]
include degeneratesurfacefixed;
\end{asy}
\begin{center}
(f)
\end{center}
\end{minipage}
\caption{Degeneracy in a Coons patch.
The dots indicate corner control points (nodes) and the open circles indicate
the points of greatest degeneracy on the boundary, as determined by the
quartic root solver:
(a) overlapping isoline mesh;
(b) overlapping patch;
(c) nonoverlapping subpatches;
(d) internally degenerate isoline mesh;
(e) internally degenerate patch;
(f) nondegenerate subpatches.}
\label{degenerate}
\end{center}
\end{figure}

The resulting mapping $\v\sigma(u,v)$ need not be bijective
\cite{Randrianarivony04,Wang05,Lin07}, even if the corner control
points form a convex quadrilateral (despite the fact that a Coons patch for
a convex polygon is always nondegenerate).
In terms of the 2D scalar cross product $\vp \cross \vq=
p_xq_y-p_yq_x$, the Coons patch is seen to be a diffeomorphism of 
the unit square $D=[0,1]\times[0,1]$ if and only if the Jacobian 
$$
J(u,v)=\Jacobian{x,y}{u,v}
=\grad_u x\cross \grad_v y=
\sum_{i,j,k,\ell=0}^3 B_i'(u)B_j(v) B_k(u)B_\ell'(v) \vP_{ij}\cross \vP_{k\ell}
$$
(the $z$ component of the corresponding 3D normal vector) is
sign-definite. Since $J(u,v)$ is a continuous function of its arguments,
this means that $J$ must not vanish anywhere on $D$. A sign reversal of the Jacobian can manifest itself as an outright overlap of the region bounded
by the curve or as an internal multivalued wrinkle, as illustrated in
Figure~\ref{degenerate}. Rendering problems, such as the black smudges
visible in Figures~\ref{degenerate}(b) and (e), can occur where isolines
collide.

Randrianarivony and Brunnett \cite{Randrianarivony04} (and
later H.~Lin \etal\ \cite{Lin07}) describe sufficient
conditions for $J(u,v)$ to be nonzero throughout $D$. In the case of a
cubic B\'ezier patch, the $36$ quantities
$$
T_{pq}=\sum_{i+k=p}\ \sum_{j+\ell=q} \vU_{i,j}\cross \vV_{k,\ell} 
{2\choose i}{3\choose k}{3\choose j}{2\choose \ell}\qquad p,q=0,1,\ldots,5,
$$
where $\vU_{i,j}=\vP_{i+1,j}-\vP_{i,j}$ and $\vV_{i,j}=\vP_{i,j+1}-\vP_{i,j}$,
are required to be of the same sign. This follows from the fact that
$J(u,v)=\sum_{p,q=0}^5 T_{pq} u^pv^q(1-u)^{5-p}(1-v)^{5-q}$.

Randrianarivony \etal\ show further that every degenerate
Coons patch can be decomposed into a finite union of nondegenerate
subpatches (some with reversed orientation). However, the adaptive subdivision
algorithm they propose to exploit this fact does not prescribe an optimal
boundary point at which to do the splitting. A better algorithm is based on the
following elementary theorem, which provides a practical means of detecting
Coons patches with degenerate boundaries.

\begin{theorem}[Nondegenerate Boundary]\label{nondegenerateboundary}
Consider a closed counter-clockwise oriented four-segment curve $p$ in the
$x$--$y$ plane such that the interior angles formed by the incoming and
outgoing tangent vectors at each node are less than or equal to $180^\degrees$.
Let $J(u,v)$ be the Jacobian of the corresponding Coons patch
constructed from $p$, with control points $\vP_{ij}$, and define the fifth-degree
polynomial
$$
f(u)=\sum_{i,j=0}^3 B_i'(u) B_j(u) \vP_{i,0}\cross (\vP_{j,1}-\vP_{j,0}).
$$
If $f(u) \ge 0$ whenever $f'(u)=0$ on $u\in(0,1)$,
then $J(u,0) \ge 0$ on $[0,1]$. Otherwise, the minimum value of $J(u,0)$
occurs at a point where $f'(u)=0$.
\end{theorem}
\begin{proof}
First we note, since $B_1'(0)=-B_0'(0)=3$ and $B_2'(0)=B_3'(0)=0$, that
$J(u,0)=3f(u)$ and
$$
J(0,0)=3f(0)=9(\vP_{1,0}-\vP_{0,0})\cross (\vP_{0,1}-\vP_{0,0}) \ge 0
$$
since this is the cross product of the outgoing tangent vectors at
$\vP_{0,0}$. Likewise, $J(1,0)=3f(1) \ge 0$.
We know that the continuous function $f$ must achieve its minimum value on
$[0,1]$ at some $u\in [0,1]$. If~$f$ were negative somewhere in $(0,1)$ we
could conclude that $f(u) < 0$, so that $u\in(0,1)$, and hence $f$ would
have an interior local minimum at $u$, with $f'(u)=0$. But this is a
contradiction, given that $f(u) \ge 0$ whenever $f'(u)=0$.
\end{proof}

The significance of Theorem~\ref{nondegenerateboundary}
is that it affords a means of detecting a point $u$ on the boundary where
the Jacobian is most negative. This requires finding roots of the
quartic polynomial
$$
f'(u)=[B_i''(u)B_j(u)+B_i'(u)B_j'(u)]\vP_{i,0}\cross (\vP_{j,1}-\vP_{j,0}).
$$
The coefficients of this quartic polynomial can be computed using the
polynomials $M_{ij}=(B_i''B_j+B_i'B_j')/3$ tabulated in Table~\ref{Mij}.
The method of Neumark \cite{Neumark65}, which relies on numerically robust
cubic and
quadratic root solvers, is then used to find algebraically all real roots
of the quartic equation $f'(u)=0$ that lie in $(0,1)$. The Jacobian
is computed at each of these points; if it is negative anywhere, the point
where it is most negative is determined. The patch is then split along an
interior line segment perpendicular to the tangent vector at this
point. The next intersection point of the patch boundary with this line
is used to split the patch into two pieces. Each of these pieces is then
treated recursively (beginning with an additional call to {\tt bezulate}, 
should the new boundary curve happen to have five segments). 

\begin{table}[tp]
\scriptsize
$$
\left(\matrix{
5-20u+30u^2-20u^3+5u^4&-3+24u-54u^2+48u^3-15u^4&-6u+27u^2-36u^3+15u^4&
-3u^2+8u^3-5u^4\cr
&\cr
-7+36u-66u^2+52u^3-15u^4&3-36u+108u^2-120u^3+45u^4&6u-45u^2+84u^3-45u^4&
3u^2-16u^3+15u^4\cr
&\cr
2-18u+45u^2-44u^3+15u^4&12u-63u^2+96u^3-45u^4&18u^2-60u^3+45u^4&8u^3-15u^4\cr
&\cr
2u-9u^2+12u^3-5u^4&9u^2-24u^3+15u^4&12u^3-15u^4&5u^4\cr
}\right)
$$
\caption{Coefficients of the polynomials $M_{ij}=(B_i''B_j+B_i'B_j')/3$.}
\label{Mij}
\end{table}

If a patch possesses only internal degeneracies, like the one in
Figure~\ref{degenerate}(d), the patch boundary is arbitrarily split into
two closed curves, say along the perpendicular to the midpoint of some
nonstraight side. The blue lines in Figures~\ref{degenerate}(b) and~(f)
illustrate such a midpoint splitting.
The arguments of Randrianarivony \etal\ \cite{Randrianarivony04} establish that
only a finite number of such subdivisions will be required to obtain
a nondegenerate patch. Nondegenerate subpatches oriented in the direction
opposite to the normal vector corresponding to the original oriented curve
should be discarded to avoid rendering interference with correctly aligned
overlying subpatches.

The blue lines in Figure~\ref{degenerate}(c) show that our quartic
algorithm generates six subpatches, a substantial improvement over the nine
subpatches produced by adaptive midpoint subdivision
\cite{Randrianarivony04} in Figure~\ref{degenerate}(b).
Figure~\ref{degenerate}(c) also emphasizes the ability of the quartic root
algorithm to detect the optimal (most degenerate) points (circled) for
splitting the boundary curve.
As mentioned earlier, in both cases, it is possible that
splitting can lead to curves with five segments. Such curves are split further
by the {\tt bezulate} algorithm so that any degeneracy of the resulting
subpatches can be addressed.

Since an algebraic quartic root solver is an explicit algorithm,
optimal subdivision of patches introduces minimal overhead compared to
adaptive midpoint subdivision. In our implementation, the costs 
of adaptive midpoint subdivision for Figures~\ref{degenerate}(b) and
Figure~\ref{degenerate}(f) were approximately the same. 
Using optimal subdivision in Figure~\ref{degenerate}(c) was $34\%$ faster
than adaptive midpoint splitting, whereas there was only $2\%$ additional
overhead in checking for boundary degeneracy in Figure~\ref{degenerate}(f)
(which possesses only internal degeneracy). Patches having only internal
degeneracy arise relatively rarely in practice, but when they do, the subpatches
obtained by adaptive midpoint subdivision also tend to exhibit internal
degeneracy. Once internal degeneracy has been detected in a patch, we find
that it is typically more efficient not to check its degenerate subpatches
for boundary degeneracy (otherwise the overhead in checking for boundary
degeneracy in Figure~\ref{degenerate}(f) would grow to $50\%$). Of course,
since our interest is not in real-time rendering but in surface
generation, the real advantage of optimal subdivision is that it can
significantly reduce the number of generated patches
(e.g.\ Figure~\ref{degenerate}(c) has one-third fewer patches than
Figure~\ref{degenerate}(b)).

\section{\texorpdfstring{An Optimized B\'ezier Inside--Outside Test}{An Optimized B\'ezier Inside-Outside Test}}\label{insidedness}
Although \PS\ has an {\tt infill\/} function for testing whether a
particular point would be painted by the \PS\ {\tt fill\/} command,
this is only an approximate digitized test corresponding to the
resolution of the output device. Our {\tt bezulate} routine
requires a vector graphics algorithm, one that yields the winding number of
an arbitrary closed piecewise B\'ezier curve about a given point. 

A straightforward generalization of the standard ray-to-infinity method for
computing winding numbers of a polygon about a point requires the
solution of a cubic equation. As is well known, the latter problem
can become numerically unstable as two or three roots begin to
coalesce. While a conventional ray-curve (or ray-patch) intersection
algorithm based on recursive subdivision \cite{Nishita90} could be employed
to count intersections by actually finding them, this typically entails
excessive subdivision.

A more efficient but still robust subdivision method for computing the
winding number of a closed B\'ezier curve arises from the topological
observation that if a point~$z$ lies outside the convex hull of a B\'ezier
segment, the segment can be continuously deformed to a straight line
segment between its endpoints, without changing its orientation relative to
the point $z$.
A given point will typically lie outside the convex hull of most segments
of a B\'ezier curve. The orientation of these segments relative to the
given point can be quickly and robustly determined, just as in the usual
ray method for polygons, to determine the contribution, if any, to the
winding number. For this purpose, Jonathan Shewchuk's public-domain
adaptive precision predicates for computational geometry \cite{shewchuk97a}
are highly recommended.

\begin{figure}[tp]
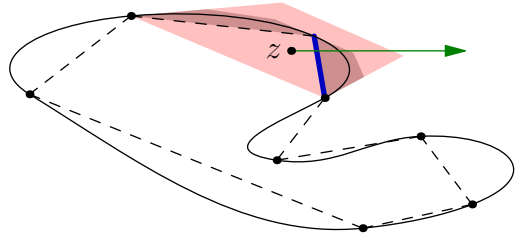

\begin{center}
\begin{asy}[width=0.49*\the\linewidth]
include inside;
\end{asy}
\caption{The {\tt B\'ezierWindingNumber} algorithm. 
Since $z$ lies inside the convex hull of one B\'ezier segment, indicated by the
light shaded region, that segment must be subdivided.
On subdivision, $z$ now lies outside the convex hulls of the subsegments,
indicated by the dark shaded regions; these subsegments may be
continuously deformed to straight line segments between their endpoints,
without crossing $z$. The usual polygon inside--outside test may then be
applied: the green ray establishes a winding number contribution of $+1$
due to the orientation of $z$ with respect to the blue line.}\label{inside}
\end{center}
\end{figure}

In the infrequent case where $z$ lies on or inside the convex hull of a
segment, de Casteljau subdivision is used to
split the B\'ezier segment about its parametric midpoint.
Typically the convex hulls of the resulting
subsegments will overlap only at their common control point, so that $z$
can lie strictly inside at most one of these hulls. This observation is
responsible for the efficiency of the algorithm: one continues subdividing
until the point is outside the convex hull of both segments or until
machine precision is reached, as illustrated in Figure~\ref{inside}.

The orientation of segments whose convex hulls do not contain $z$ can be
handled by using the topological deformation property together with
adaptive precision predicates.  Denoting by 
{\tt straightContribution(\sf{P,Q,z})} the usual ray method for
determining the winding number contribution of a line segment $\overline{PQ}$
relative to a point {\sf z}, the contribution from a B\'ezier segment~{\sf S}
can be computed as {\tt curvedContribution(\sf{S,z})}
(Algorithm~\ref{curvedcontribution}).

\begin{algorithm}[H]
  \SetKwData{w}{W}
  \SetKwData{s}{S}
  \SetKwData{sss}{s}
  \SetKwData{z}{z}
  \KwIn{segment \s, pair \z}
  \KwOut{winding number contribution of \s about \z}
  \w $\leftarrow$ $0$;\\
  \eIf{\z lies within or on the convex hull of \s} {
    \ForEach{subsegment \sss of \s}{
      \w $\leftarrow$ \w + {\tt curvedContribution}(\sss,\z)\;
    }
  }
  {
    \w $\leftarrow$ \w + {\tt straightContribution}(\s.beginpoint,\s.endpoint,\z)\;
  }
  \Return \w;
\caption{{\tt curvedContribution({\sf S,z})} determines the winding number
contribution from a B\'ezier segment~{\sf S} about {\sf z}.}\label{curvedcontribution}
\end{algorithm}

The winding number for a closed curve $p$ about $z$ may then be evaluated
with the algorithm
{\tt b\'ezierWindingNumber({\sf C,z})} (Algorithm~\ref{bezierwindingnumber}).

\begin{algorithm}[H]
  \SetKwData{w}{W}
  \SetKwData{p}{C}
  \SetKwData{s}{S}
  \SetKwData{z}{z}
  \KwIn{curve \p, pair \z}
  \KwOut{winding number of \p about \z}
  \w $\leftarrow$ $0$;\\
  \ForEach{segment \s of \p}{
    \eIf{\s is straight}{
      \w $\leftarrow$ \w + {\tt straightContribution}(\s.beginpoint,\s.endpoint,\z)\;
    }
    {
      \w $\leftarrow$ \w + {\tt curvedContribution}(\s,\z)\;
    }
  }
  \Return \w;
\caption{{\tt b\'ezierWindingNumber({\sf C,z})} computes the winding number
of a closed B\'ezier curve {\sf C} about {\sf z}.}\label{bezierwindingnumber}
\end{algorithm}

A practical simplification of the above algorithm is the widely used
optimization of testing whether a point is inside the 2D bounding box of
the control points rather than their convex hull.
Since the convex hull of a B\'ezier segment is contained within
the bounding box of its control points, one can replace ``convex hull'' by
``control point bounding box'' in the above algorithm without modifying its
correctness. One can easily check numerically that the cost of the additional
spurious subdivisons is well offset by the computational savings in testing
against the control point bounding box.

\section{Global Bounds of Directionally Monotonic Functions}\label{monotonic}
\begin{figure}[tp]
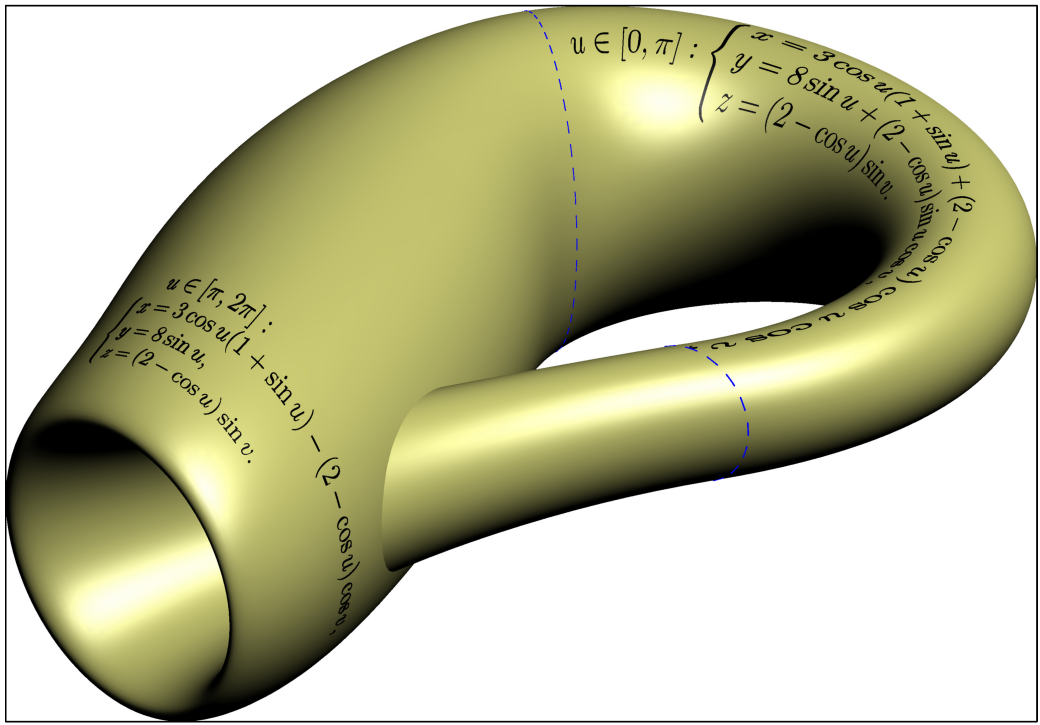

\begin{asy}[inline,width=\the\linewidth]
include Klein;
\end{asy}
\caption{A B\'ezier approximation to a projection of a four-dimensional
Klein bottle to three dimensions. The {\tt FunctionMax} algorithm was
used to determine the optimal field of view for this symmetric perspective
projection of the scene from the camera location $(25.09,-30.33,19.37)$ 
looking at $(-0.59,0.69,-0.63)$. The extruded 3D \TeX\ equations embedded
onto the surface provide a parametrization for the surface over the domain
$u\times v\in[0,2\pi]\times[0,2\pi]$.}
\label{klein}
\end{figure}

We now present efficient algorithms for computing global bounds of
real-valued directionally monotonic functions $f:\R^3\to \R$
defined over a B\'ezier surface $\v\sigma(u,v)$.
By {\it directionally monotonic} we mean that the restriction of $f$ to
each of the three Cartesian directions is a monotonic function; if $f$ is 
differentiable this means that $f$ has sign-semidefinite partial derivatives.
These algorithms can be used to compute the 3D bounding box of a B\'ezier
surface, the bounding box of its 2D projection, or the optimal
field-of-view angle for sizing a 3D scene (cf.~Fig.~\ref{klein}).
The key observation is that the convex hull property of a B\'ezier patch holds
independently in each direction and even under inversions like $z\to 1/z$.

A na\"ive approach to computing the bounding box of a B\'ezier patch
requires subdivision whenever the 3D bounding
boxes overlap in any of the three Cartesian directions. 
However, the number of required subdivisons can be greatly reduced by
decoupling the three directions: in Algorithm~\ref{CartesianMax},
the problem is split into finding
the maximum and minimum of the three {\it Cartesian axis projections\/}
$f(x,y,z)=x$, $f(x,y,z)=y$, and $f(x,y,z)=z$ evaluated over the patch.
This requires a total of six applications of Algorithm~\ref{CartesianMax}.
By convexity, the extrema of these special choices
for $f$ over a convex polyhedron $\cal C$ occur at vertices of $\cal C$.

More general choices of directionally monotonic functions $f$ are also of
interest. For example, to determine the bounding box of the 2D perspective
projection (based on similar triangles) of a surface, one can apply
Algorithm~\ref{FunctionMax} in eye coordinates to the functions
$f(x,y,z)=x/z$ and $f(x,y,z)=y/z$.
This is useful for sizing a 3D object in terms of its 2D projection.
For example, these functions were used to calculate the optimal
field-of-view angle $13.4^\degrees$ for the Klein bottle shown in
Figure~\ref{klein}.

For an arbitrary directionally monotonic function $f$, we note that
\begin{equation}
\v\sigma \subset {\cal C}\Rightarrow f(\v\sigma)\subset f({\cal C}).
\label{convex}
\end{equation}

Our algorithms exploit Eq.~(\ref{convex}) together with de Casteljau's
subdivision algorithm and the fact that a B\'ezier 
patch is confined to the convex hull of its 
control points.  However, a patch is only guaranteed to intersect its
convex hull at the four corner nodes.

For the special case where $f$ is a projection onto the Cartesian axes,
the function {\tt CartesianMax}$(f,\vP,f(\vP_{00}),d)$ given in
Algorithm~\ref{CartesianMax} computes the global maximum $M$ of 
a Cartesian axis projection $f:\R^3\to \R$ over a B\'ezier patch $\vP$ to
recursion depth~$d$. Here, the value $f(\vP_{00})$ provides a
convenient starting value (lower bound) for $M$; if the maximum of a
surface consisting of several patches is desired, the value of $M$ from
previous patches is used to seed the calculation for the subsequent one.
The algorithm exploits the fact that the extrema of each
coordinate over the convex hull~$\cal C$ of~$\vP$ occur
at vertices of $\cal C$.
First, one replaces~$M$ by the maximum of $f$ evaluated at the four corner nodes
and the previous value of~$M$. If the maximum of the function evaluated
at the remaining $12$ control points is less than or equal to~$M$, the subpatch
can be discarded (by Eq.~\ref{convex}, noting that the maximum of
$f({\cal C})$ occurs at a control point and hence cannot exceed $M$).
Otherwise, the patch is subdivided along the
$u=v=1/2$ isolines and the process is repeated using the new value of~$M$.
The method quickly converges to the global maximum of~$f$ over
the entire patch.

\begin{algorithm}[htp]
  \KwIn{real function {\sf f(triple)}, patch {\sf $\vP$}, real {\sf M},
    integer {\sf depth}}
  \KwOut{real {\sf M}}
  \SetKwData{f}{f}
  \SetKwData{p}{$\vP$}
  \SetKwData{s}{S}
  \SetKwData{mm}{M}
  \SetKwData{VV}{V}
  \SetKwData{Depth}{depth}
  \SetKwData{Or}{{\bf or}}
  $\mm \leftarrow \max(\mm,\f(\p_{00}),\f(\p_{03}),\f(\p_{30}),\f(\p_{33}))$\;
  \If{$\Depth = 0$}{\Return \mm\;}
  $\VV\leftarrow\max(\f(\p_{01}),\f(\p_{02}),\f(\p_{10}),\f(\p_{11}),\f(\p_{12}),
  \f(\p_{13}),\linebreak\phantom{\VV\leftarrow\max(}
  \f(\p_{20}),\f(\p_{21}),\f(\p_{22}),\f(\p_{23}),\f(\p_{31}),\f(\p_{32}))$\;
  \If{$\VV \le \mm$}{\Return \mm\;}
  \ForEach{subpatch \s of \p}{
    $\mm \leftarrow \max(\mm,{\tt FunctionMax}(\f,\s,\mm,\Depth-1))$\;
  }
  \Return \mm\;
  \caption{{\tt CartesianMax({\sf f},$\vP$,{\sf M},{\sf depth})} returns the
    maximum of {\sf M} and the global bound of a Cartesian component~{\sf
      f} of a B\'ezier patch {$\vP$} evaluated to recursion level {\sf depth}.}
  \label{CartesianMax}
\end{algorithm}

For a general directionally monotonic function $f$
(consider $f(x,y,z)=xy$ over ${\cal C}=\partial\{(x,y,0) : 0 \le x \le
1,\ 0\le y \le x\}$), the maximum of $f({\cal C})$ need not occur at
vertices of $\cal C$: one instead needs to examine the function value at
the appropriate vertex of the bounding box of ${\cal C}$. For
example, if $f$ is a monotonic increasing function in each of the
three Cartesian directions,
\begin{equation}
{\cal C}\subset\Box(\va,\vb) \Rightarrow f({\cal C})\subset [f(\va),f(\vb)],
\label{directionallymonotonic}
\end{equation}
where $\Box(\va,\vb)$ denotes the 3D box with minimal and maximal vertices
$\va$ and $\vb$, respectively.

The global maximum $M$ of a
directionally monotonic increasing function $f:\R^3\to \R$
over a B\'ezier patch~$\vP$ can then be efficiently computed to recursion
depth~$d$ by calling the function {\tt FunctionMax}$(f,\vP,f(\vP_{00}),d)$
given in Algorithm~\ref{FunctionMax}.
First, one replaces~$M$ by the maximum of $f$ evaluated at the four corner nodes
and the previous value of $M$. One then computes the vertex $\vb$ of the
bounding box of the convex hull ${\cal C}$ of $\vP$.
If the maximum of the function evaluated at $\vb$ is less than or equal to
$M$, the subpatch can be discarded.  Otherwise, the patch is subdivided
along the $u=v=1/2$ isolines and the process is repeated using the new
value of $M$.

\begin{algorithm}[htp]
  \KwIn{real function {\sf f(triple)}, patch {\sf $\vP$}, real {\sf M},
    integer {\sf depth}}
  \KwOut{real {\sf M}}
  \SetKwData{f}{f}
  \SetKwData{p}{$\vP$}
  \SetKwData{s}{S}
  \SetKwData{mm}{M}
  \SetKwData{x}{x}
  \SetKwData{y}{y}
  \SetKwData{z}{z}
  \SetKwData{Depth}{depth}
  $\mm \leftarrow \max(\mm,\f(\p_{00}),\f(\p_{03}),\f(\p_{30}),\f(\p_{33}))$\;
  \If{\Depth = 0}{\Return \mm\;}
  $\x\leftarrow \max(\hat {\v x}\cdot \p_{ij}: 0 \le i,j \le 3)$\;
  $\y\leftarrow \max(\hat {\v y}\cdot \p_{ij}: 0 \le i,j \le 3)$\;
  $\z\leftarrow \max(\hat {\v z}\cdot \p_{ij}: 0 \le i,j \le 3)$\;
  \If{$f((\x,\y,\z)) \le \mm$}{
    \Return \mm\;}
  \ForEach{subpatch \s of \p}{
    $\mm \leftarrow \max(\mm,{\tt FunctionMax}(\f,\s,\mm,\Depth-1))$\;
  }
  \Return \mm\;
  \caption{{\tt FunctionMax({\sf f,$\vP$,M,depth})} returns the maximum
    of {\sf M} and the global bound of a real-valued directionally
    monotonic increasing function~{\sf f} over a B\'ezier patch {\sf $\vP$}
    evaluated to recursion level {\sf depth}. Here $\hat {\v x}$,
    $\hat {\v y}$, $\hat {\v z}$ are the Cartesian unit vectors.}
  \label{FunctionMax}
\end{algorithm}

\section{3D Vector Typography}\label{Asymptote}
Donald Knuth's \TeX\ system \cite{Knuth86}, the de-facto standard for
typesetting mathematics, uses B\'ezier curves to represent 2D characters.
\TeX\ provides a portable interface that yields consistent, publication quality
typesetting of equations, using subtle spacing rules derived from
centuries of professional mathematical typographical experience. 
However, while it is often desirable to illustrate abstract mathematical
concepts in \TeX\ documents, no compatible descriptive standard
for technical mathematical drawing has yet emerged.

The recently developed \Asymptote\ language\footnote{available from
\url{http://asymptote.sourceforge.net} under the GNU Lesser General
Public License.} aims to fill this gap by providing a portable
\TeX-aware tool for producing 2D and 3D vector graphics \cite{Bowman08}.
In mathematical applications, it is important to typeset labels and
equations with \TeX\ for overall consistency between the text and graphical
elements of a document. In addition to providing access to the
\TeX\ typesetting system in a 3D context, \Asymptote\ also fills in a gap for
nonmathematical applications. While open source 3D
bit-mapped text fonts are widely available,\footnote{For example, see
  \url{http://www.opengl.org/resources/features/fontsurvey/}.}
resources currently available for scalable (vector) fonts appear to be
quite limited in three dimensions.

\Asymptote\ was inspired by John Hobby's \MP\ (a
modified version of \MF, the program that Knuth wrote to generate the 
\TeX\ fonts), but is more powerful, has a cleaner syntax, and uses \acro{IEEE}
floating point numerics.  An important feature of
\Asymptote\ is its use of the simplex linear programming method to solve overall
size constraint inequalities between fixed-sized objects (labels, dots, and
arrowheads) and scalable objects (curves and surfaces). This means that the user
does not have to scale manually the various components of a figure by
trial-and-error. The 3D versions of \Asymptote's deferred drawing routines
rely on the efficient algorithms for computing the bounding
box of a B\'ezier surface, along with the bounding box of its 2D projection, 
described in Sec.~\ref{monotonic}. \Asymptote\ natively
generates \PS, \PDF, \SVG, and \PRC\ \cite{PRCFormat08} vector
graphics output. The latter is a highly compressed 3D format that is typically
embedded within a \PDF\ file and viewed with the widely available
{\sc Adobe Reader} software.

The biggest obstacle that was encountered in generalizing \Asymptote\ to produce
3D interactive output was the fact that \TeX\ is fundamentally a 2D program.
In this work, we have developed a technique for embedding 2D vector
descriptions, like \TeX\ fonts, as 3D surfaces 
(2D vector graphics representations of \TeX\ output can be extracted with
a technique like that described in Ref.~\cite{Bowman09}).
While the general problem of filling an arbitrary 3D closed
curve is ill-posed, there is no ambiguity in the important special case of
filling a planar curve with a planar surface.

Since our procedure transforms text into B\'ezier patches, which are the surface
primitives used in \Asymptote, all of the existing 3D \Asymptote\ algorithms
can be used without modification.
Together with the 3D generalization of the \MF\ curve operators
described by \cite{Bowman07,Bowman08}, these algorithms comprise the
3D foundation for the \TeX-aware vector graphics language \Asymptote.

\subsection{3D Arrowheads}\label{arrows}
\begin{figure}[tph]
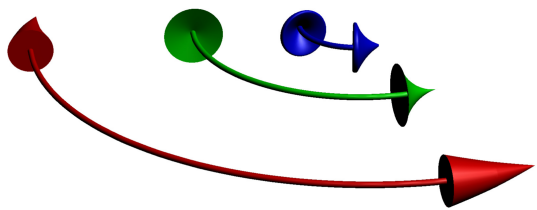

\begin{asy}[inline,width=0.49*\the\linewidth]
include arrows3a;
\end{asy}
\caption{Three-dimensional revolved arrowheads in \Asymptote.}\label{arrows3a}
\end{figure}

\begin{figure}[tph]
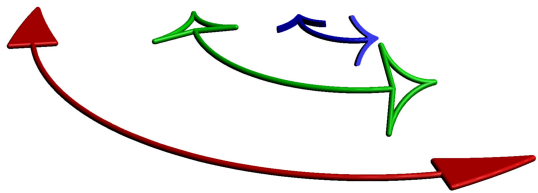

\begin{asy}[inline,width=0.49*\the\linewidth]
include arrows3b;
\end{asy}
\caption{Planar arrowheads in \Asymptote.}\label{arrows3b}
\end{figure}

Arrows are frequently used in illustrations to draw attention to important
features. We designed curved 3D arrowheads that can be viewed from a
wide range of angles. For example, the default 3D arrowhead was formed by
bending the control points of a cone around the tip of a B\'ezier curve.
Planar arrowheads derived from 2D arrowhead styles are also implemented;
they are oriented by default on a plane perpendicular to the initial
viewing direction. Examples of these arrows are displayed in
Figures~\ref{arrows3a} and~\ref{arrows3b}. The {\tt bezulate}
algorithm was used to construct the upper and lower faces of the filled
(red) planar arrowhead in Fig.~\ref{arrows3b}.

\subsection{Double Deferred Drawing}\label{doubledeferred}
Journal size constraints typically dictate the final width and height,
in \PS\ coordinates, of a 2D or projected 3D
figure. However, it is often convenient for users to work in more physically
meaningful coordinates. This requires \It{deferred drawing:} a graphical object
cannot be drawn until the actual scaling of the user coordinates (in terms
of \PS\ coordinates) is known \cite{Bowman08}. One therefore needs
to queue a function that can draw the scaled object later, when this
scaling is known. \Asymptote's high-order functions provide a
flexible mechanism that allows the user to specify either or both of the 3D
model dimensions and the final projected 2D size. This requires two levels of
deferred drawing, one that first sizes the 3D model and one that scales the
resulting picture to fit the requested 2D size \cite{Bowman09}.
The 3D bounding box of a B\'ezier surface, along with the bounding box of
its 2D projection, can be efficiently computed with the method described
in Section~\ref{monotonic}.

\subsection{Efficient Rendering}
Efficient algorithms for determining the bounding box of a
B\'ezier patch also have an important application in rendering.
Knowing the bounding box of a B\'ezier patch allows one to determine, at a
high level, whether it is in the field of view: offscreen B\'ezier patches
can be dropped before mesh generation occurs \cite{Hasselgren09}.
This is particularly important for a spatially adaptive algorithm as used in
\Asymptote's \OpenGL-based renderer,
which resolves the patch to one pixel precision at all zoom levels.
Moreover, to avoid subdivision cracks, renderers typically resolve visible
surfaces to a uniform resolution. It is therefore important that offscreen
patches do not force an overly fine mesh within the viewport.
As a result of these optimizations, the native \Asymptote\ adaptive renderer is
typically comparable in speed with the fixed-mesh \PRC\ renderer in {\sc
  Adobe Reader}, even though the former yields higher quality, true vector
graphics output.

\section{Conclusions}
In this work we have developed methods that can be used to lift
smooth fonts, such as those produced by \TeX, into 3D.
Treating 3D fonts as surfaces allows for arbitrary 3D text manipulation,
as illustrated in Figures~\ref{label3zoom} and~\ref{klein}.
The {\tt bezulate} algorithm allows one to
construct planar B\'ezier surface patches by decomposing (possibly nonsimply
connected) regions bounded by simple closed curves into
subregions bounded by closed curves with four or fewer segments.
The method relies on an optimized subdivision algorithm for testing whether a
point lies inside a closed B\'ezier curve, based on the 
topological deformation of the curve to a polygon.
We have also shown how degenerate Coons patches can be efficiently detected
and split into nondegenerate subpatches. This is required to avoid both
patch overlap at the boundaries of the underlying curve and rendering
artifacts (patchiness, smudges, or wrinkles) due to normal reversal.

We have illustrated applications of these techniques in the open
source vector graphics programming language \Asymptote, which we believe is
the first software to lift \TeX\ into 3D. This represents an important
milestone for publication-quality scientific graphing.

\appendix
\section{B\'ezier Approximation of a Sphere}\label{solids}
As previously emphasized, although conic sections (quadrics) may be
accurately represented by NURBS surfaces, the
language of high-end printers, \PS, supports only B\'ezier
curves and surfaces. Although \PS\ is only a 2D language, vector graphics
projections of B\'ezier surfaces are nevertheless possible using
tensor-product patch shading and hidden surface splitting along
approximations to the visible surface horizon.

Here we illustrate that a sphere may be approximated to high
graphical accuracy by a B\'ezier surface with only $8$ patches, one for each
octant, following a procedure suggested in Ref.~\cite{Ostby88}. The patch
describing an octant is degenerate at the pole: two of the nodes are placed
there, with the other two placed along the equator, $90^\degrees$ apart in
longitude.

Following Knuth, a unit quarter circle is approximated in \Asymptote\ 
``with less than 0.06\% error'' \cite{Knuth86b}, using the control points
$\{(1,0),(1,a),(a,1),(0,1)\}$, where $a=\frac{4}{3}(\sqrt{2}-1)$.
This value of~$a$ is determined by requiring that the
third-order B\'ezier midpoint lie on the unit circle at
$(1/\sqrt{2},1/\sqrt{2})$. (Other methods of approximating circular arcs by
B\'ezier curves have been described in Refs.~\cite{Bezier86},
\cite{Fang98}, and \cite{Piegl03circle}.)

\begin{figure}[tp]
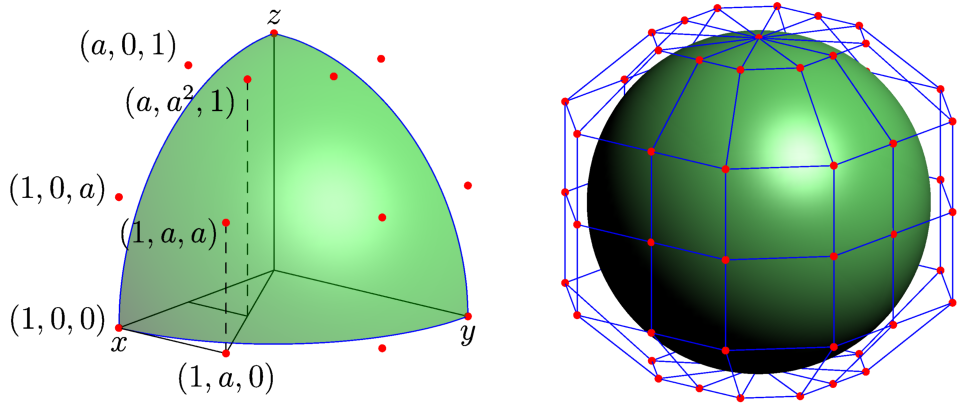

\begin{center}
\begin{minipage}{0.49\linewidth}
\begin{asy}[inline]
include unitoctant;
\end{asy}
\end{minipage}
\begin{minipage}{0.49\linewidth}
\begin{asy}[inline]
include unitsphere;
\end{asy}
\end{minipage}
\caption{B\'ezier approximation to a unit sphere. The red dots indicate
control points.}
\label{unitsphere}
\end{center}
\end{figure}

The above prescription immediately determines the three
circular arcs describing the patch boundary for a unit spherical octant.
Let us place $\vP_{00}$ at $(1,0,0)$, $\vP_{03}=\vP_{13}=\vP_{23}=\vP_{33}$
at $(0,0,1)$, and $\vP_{30}$ at $(0,1,0)$. The remaining control points
$\{\vP_{11},\vP_{12},\vP_{21},\vP_{22}\}$ are chosen to make the 
surface nearly spherical and the interface with adjacent octants
smooth (have continuous first derivatives at the patch boundaries).
The point $\vP_{11}$ is chosen (on the tangent plane at $x=1$) to be the
vector sum $\vP_{10}+\vP_{01}-\vP_{00}=(1,a,0)+(1,0,a)-(1,0,0)=(1,a,a)$.
We also require that the triangle in the $x$--$y$ plane formed by 
the origin and the projections of~$\vP_{12}$ onto the $x$--$y$ plane and
the $x$ axis is similar to the corresponding triangle for $\vP_{11}$.
This implies that $\vP_{12}=(a,a^2,1)$. Similarly, we determine
$\vP_{22}=(a^2,a,1)$ and $\vP_{21}=(a,1,a)$. The final B\'ezier patch and
resulting approximation to a unit sphere, with the control point mesh shown in
blue, are illustrated in Figure~\ref{unitsphere}.
We found numerically that the radius of this approximate sphere, 
generated with a $12\times 7$ control point mesh, varies by less than
$0.052\%$, well below the tolerance $0.1\%$ to which Figure~8 of
Ref.~\cite{piegl03} was drawn using a much finer $22\times 13$ control
point mesh.

\section*{Acknowledgments}

We thank Philippe Ivaldi, Radoslav Marinov, Malcolm Roberts, Jens Schwaiger,
and Olivier Guib\'e for discussions and assistance in implementing the
algorithms described in this work. Special thanks goes to Andy Hammerlindl,
who designed and implemented much of the underlying \Asymptote\ language,
and to Michail Vidiassov, who helped write the \PRC\ driver. Financial
support for this work was generously provided by the Natural Sciences and
Engineering Research Council of Canada.

\bibliography{tex/refs}

\end{document}